# Generating Inputs for Grammar Mining using Dynamic Symbolic Execution

Andreas Pointner[a] , Josef Pichler[b] , and Herbert Prähofer[c]

a  Advanced Information Systems and Technology, University of Applied Sciences Upper Austria, Hagenberg, Austria

b  Department for Software Engineering, University of Applied Sciences Upper Austria, Hagenberg, Austria

c  Institute for System Software, Johannes Kepler University, Linz, Austria

**Abstract**   A vast number of software systems include components that parse and process structured input. In addition to programming languages, which are analyzed by compilers or interpreters, there are numerous components that process standardized or proprietary data formats of varying complexity. Even if such components were initially developed and tested based on a specification, such as a grammar, numerous modifications and adaptations over the course of software evolution can make it impossible to precisely determine which inputs they actually accept.

In this situation, grammar mining can be used to reconstruct the specification in the form of a grammar. Established approaches already produce useful results, provided that sufficient input data is available to fully cover the input language. However, achieving this completeness is a major challenge. In practice, only input data recorded during the operation of the software systems is available. If this data is used for grammar mining, the resulting grammar reflects only the actual processed inputs but not the complete grammar of the input language accepted by the software component. As a result, edge cases or previously supported features that no longer appear in the available input data are missing from the generated grammar.

This work addresses this challenge by introducing a novel approach for the automatic generation of inputs for grammar mining. Although input generators have already been used for fuzz testing, it remains unclear whether they are also suitable for grammar miners. Building on the grammar miner Mimid, this work presents a fully automated approach to input generation. The approach leverages Dynamic Symbolic Execution (DSE) and extends it with two mechanisms to overcome the limitations of DSE regarding structured input parsers. First, the search for new inputs is guided by an iterative expansion that starts with a single-character input and gradually extends it. Second, input generation is structured into a novel three-phase approach, which separates the generation of inputs for parser functions.

The proposed method was evaluated against a diverse set of eleven benchmark applications from the existing literature. Results demonstrate that the approach achieves precision and recall for extracted grammars close to those derived from state-of-the-art grammar miners such as Mimid. Notably, it successfully uncovers subtle features and edge cases in parsers that are typically missed by such grammar miners. The effectiveness of the method is supported by empirical evidence, showing that it can achieve high performance in various domains without requiring prior input samples.

This contribution is significant for researchers and practitioners in software engineering, offering an automated, scalable, and precise solution for grammar mining. By eliminating the need for manual input generation, the approach not only reduces workload but also enhances the robustness and comprehensiveness of the extracted grammars. Following this approach, software engineers can reconstruct specification from existing (legacy) parsers.

# The Art, Science, and Engineering of Programming

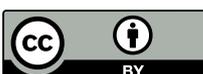

**Generating Inputs for Grammar Mining using Dynamic Symbolic Execution**

## 1 Introduction

The input specification of a program is crucial for understanding its functionality [10]. Furthermore, black-box testing of software systems relies on a defined set of inputs, which necessitates a clear and precise input specification for proper input generation.

Unfortunately, comprehensive and accurate input specifications are uncommon. If available, they might be outdated, incomplete, or incorrect [35]. Figuring out a program's input specification can be difficult, even when the source code is accessible.

A program that works with structured inputs usually has some sort of upstream parser, before executing the business logic of the program. To understand the structure of the input, it is necessary to analyze those parsers, and understand their behavior. A parser's input language can be specified by context-free grammars. Grammar mining is a technique for extracting grammars from parser-like programs. It can be done in various different ways, most of them relying on the availability of input samples. The following paragraphs provide an overview of grammar mining approaches and their basic input requirements.

There is a high variety of black-box grammar mining approaches [3, 5, 22, 36], which highly rely on the availability of input samples. Only one of them, REINAM [36], integrates dynamic symbolic execution for finding a few initial input samples and then continues with reinforcement learning. However, the quality of the result depends mainly on reinforcement learning process, which is likely to ignore details hidden in the parser implementation.

White-box grammar mining approaches [17, 18] also rely on input samples. While Autogram [18] is based on tracing the data flow of input characters using dynamic tainting, Mimid [17] traces the dynamic control flow to build parse trees, which then can be transformed into the resulting grammar. Both approaches do have some generalization steps, which reduce the need for a high variety of inputs, but still require having all possible structures present in the inputs.

Sochor et al. [32, 33] showed promising results with a gray-box approach. They no longer need a large amount of input samples, but just a few seed inputs. The approach generates an initial grammar from the seed inputs which is then refined with a grammar learning algorithm.

All the presented approaches depend on the availability of a set of input samples. Consequently, the quality of the inputs plays a critical role in the results of the grammar miner. If certain structures are missing from the inputs, the grammar mining process will likely overlook them as well. Furthermore, collecting and preparing the inputs can be both time consuming and error prone, as it not only requires considerable effort but also a deep understanding of the system.

In this work, we thus present a grammar mining approach that eliminates the need for handwritten inputs. We leverage the grammar miner Mimid [17] and develop a specialized input generator tailored for the grammar mining process. The core concept is to utilize dynamic symbolic execution for the successive expansion of inputs, starting from a single character. Additionally, to overcome the path explosion problem [8], we introduce a three-phase approach that generates inputs for parser functions separately.

This raises the following research questions, which we address in this work:

RQ 1: Can our extended dynamic symbolic execution approach generate inputs that accurately reflect the structural patterns recognized by the parser for effective grammar mining?

RQ 2: How does the quality of the grammars mined by the inputs generated by our approach compare to the grammars minded by other automated input generation approaches?

The remainder of this work is structured as follows: In Section 2, we present the running example, provide an overview of Mimid [17], and explain dynamic symbolic execution (DSE). Section 3 describes our approach to generate inputs for grammar mining. Our research method, evaluation examples, and evaluation methodology are described in Section 4. The results of our experiments are presented in Section 5. In Section 6, we analyze our findings and discuss potential limitations of our work. A comparison with related work is included in Section 7. Finally, Section 8 concludes the work with some final remarks and an outlook on future work.

## 2 Background

This section introduces the running examples as well as the background work on which our approach is based. First, we introduce a running example to explain the Mimid grammar miner, which we later reuse to illustrate our approach. Then we continue with a description of the grammar miner Mimid, which serves as the foundation for our method. Additionally, the concepts of dynamic symbolic execution and fuzzing for highly structured inputs are discussed.

### 2.1 Running Example

As a running example for explaining Mimid and our approach in subsequent sections, we introduce the Microjson parser program for parsing JSON text taken from [28]. Listing 1 shows parts of the Microjson program. It defines a set of functions, such as *json_raw*, *json_dict*, *json_list*, *json_string*, *json_number* or *decode_escape*. Those functions receive the input stream *stm* as input.

In the Microjson program, *json_raw* is the entry point for the parser. It is designed as a dispatch method, checking which character the sequence starts with, parsing it, and then calling the appropriate function, e.g., calling function *json_dict* when an opening brace '{' has been detected.

The function *json_dict* shows how a typical parsing function is implemented. A JSON dictionary starts with an opening brace '{' and ends with a closing brace '}' character. As the opening brace has already been parsed by the *json_raw* method, this starting character is skipped by calling *stm.next()*. The *result* variable will collect the JSON dictionary entries. Variable *expect_key* holds the state deciding if another dictionary entry has to follow. The method then continues with a loop, where the next character is investigated. In an if-statement the different cases are distinguished:

Generating Inputs for Grammar Mining using Dynamic Symbolic Execution

First, a closing brace '}' terminates the method and returns the result. With a comma ',', the *expect_key* state is set to value '2', meaning that another dictionary entry has to follow. This state is checked in line 21 before returning the result. It results in an error because an entry is expected but not found. This also leads to the fact that the parser allows multiple commas (',') between entries. With a double quote '"' parsing of a new dictionary entry is started. An entry consists of a *json_string* call, followed by a colon ':' and a *json_raw* call. In all other cases, an error is raised.

The function *decode_escape* parses escape sequences in JSON strings. It tries to retrieve the decoded string from ESC_MAP. If it exists, it returns the decoded string. Otherwise, it checks for the character 'u', to detect a Unicode character. If that is not the case, the input character is accepted and returned. Note, that by this implementation every character except 'u' is accepted.

## 2.2 Mimid

Mimid is a grammar mining approach presented in Gopinath et al. [17]. It showed promising results in extracting human-readable grammars from C++ and Python programs.

Mimid works by executing and observing the program execution for multiple handcrafted inputs to identify which character has been accessed in which control flow node. The approach is based on the idea that in a recursive descent parser, alternative rules are evaluated until the token is parsed correctly. After that, the parser does not look back at that character anymore. Such control flow nodes can then either be function calls, or pseudo-function calls such as loops and conditionals. Based on these control flow nodes, they are able to derive parse trees. After that, they process parse trees to derive the grammar. Thus, the resulting grammar is derived from the original functions that were responsible for parsing the input.

Figure 1 shows how the program from Listing 1 analyzes a given input string. On the left-hand side, an excerpt of the parse tree is shown, while on the right-hand side the functions/pseudo-functions that are responsible for parsing which part of the input are shown. The example input '{"abc":123,"xyz":[]}' is first analyzed by the *json_raw* method. As it starts with a '{' it is further processed by the *json_dict* function. Then the dictionary contains two entries, each of them consisting of a *json_string* followed by a colon and again followed by another *json_raw* element. Finally, the input is closed by a '}' character. This results in the parse tree shown on the left side of the figure, which can then be derived into grammar rules as shown in Listing 2.

**Requirements for Inputs and Results of Mimid**   After Mimid has extracted an initial grammar, a generalization of tokens as well as a detection of repetitions is applied. During the generalization of the tokens, Mimid tries to widen characters into generic groups, such as digits, alphanumeric characters, or any characters. This allows to create a more generalized and more concise grammar. For the approach presented in this work, it means that it is not necessary to generate every possible input at every position, as this will be found out in this generalization step. In addition, Mimid detects repeating patterns, by using an adapted version of the prefix tree acceptor

▪ **Listing 1** Parts of the Microjson [28] parser that are used as a running example in this work (simplified).

```
 1  def json_raw(stm):
 2      ...
 3      c = stm.peek()
 4      if c == '"': return json_string(stm)
 5      elif c == '{': return json_dict(stm)
 6      elif c == '[': return json_list(stm)
 7      ...
 8      elif c in NUMSTART: return json_number(stm)
 9      else: raise JSONError(...)
10  def json_list(stm): ...
11  def json_string(stm): ...
12  def json_number(stm): ...
13  def json_dict(stm):
14      stm.next() # skip over '{'
15      result = {}
16      expect_key = 1
17      while True:
18          stm.skipspaces()
19          c = stm.peek()
20          if c == '}':
21              if expect_key == 2:
22                  raise JSONError(E_TRUNC, stm, pos)
23              stm.next()
24              return result
25          elif c == ',':
26              stm.next()
27              expect_key = 2
28              continue
29          elif c == '"':
30              ...
31              key = json_string(stm)
32              stm.skipspaces()
33              c = stm.next()
34              if c != ':':
35                  raise JSONError(...)
36              stm.skipspaces()
37              val = json_raw(stm)
38              ...
39              expect_key = 0
40          raise JSONError(...)
41  def decode_escape(c, stm):
42      v = ESC_MAP.get(c, None)
43      if v is not None: return v
44      elif c != 'u': return c
45      # decode unicode escape \u1234 ...
```

**Generating Inputs for Grammar Mining using Dynamic Symbolic Execution**

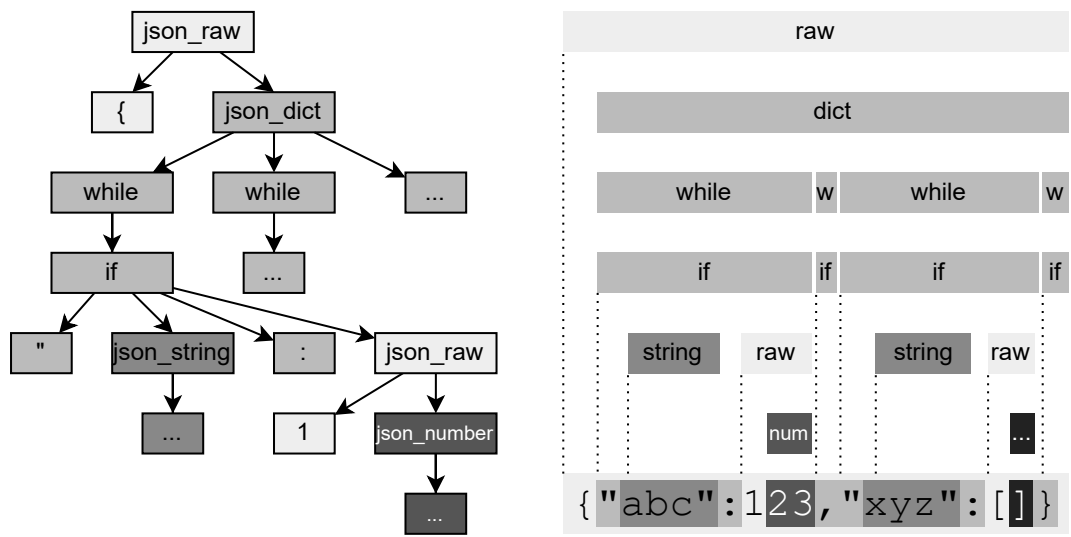

- **Figure 1** The parsing of inputs shown for the Microjson [28] program. The tree on the left shows a simplified parse tree. The right-hand side shows the inputs, and which parts of the inputs are parsed by which function or pseudo-function.

- **Listing 2** Grammar rules *json_raw*, *json_dict* and *decode_escape* of the JSON grammar extracted by Mimid [17].

```
1  …
2  <json_raw> ::= '"' <json_string> | '{' <json_dict> | …
3  <json_dict> ::= '}' | ('"' <json_string> ':' <json_raw> ',' )* '"' <json_string> ':' <json_raw> '}'
4  …
5  <decode_escape> ::= '"' |'/' |'b' |'f' |'n' |'r' |'t'
6  …
```

algorithm [11]. Experiments showed that even with just having two repetitions, the algorithm is able to correctly detect repetitions. This again means for the input generation that it is not necessary to generate a high amount of these repeating patterns, but just two.

The JSON grammar (cf. Listing 2) that Gopinath et al. [17] extracted from the Microjson [28] parser shows that the Mimid extraction process depends on the available inputs. The previously described *json_dict* function in Listing 1 allows multiple commas between dictionary entries. The sample inputs that are used in the evaluation of Mimid do not include that feature. This leads to the fact that the resulting grammar also does not contain repetitions of commas.

Another example, which shows that the inputs affect the final grammar (cf. Listing 2), can be seen in the *decode_escape* method shown in Listing 1. This method shows that every character except 'u' will be accepted by the parser (cf. line 44). The *decode_escape* rule from the Mimid paper does not show that either.

These two examples show that the results of the grammar mining approach are highly dependent on the quality and variety of the inputs. The JSON grammar, presented in the Mimid paper, shows a very accurate representation of the *JSON* *specification*. Nevertheless, it is not completely accurate regarding the *parser implementation*. We hypothesize that with an automated input generation approach such corner cases are more likely to be detected, thus allowing the generation of a more accurate grammar of the language that the program actually accepts.

## 2.3 Dynamic Symbolic Execution

While the key idea behind symbolic execution was introduced nearly four decades ago [9, 20], it only recently became practical. This advancement is due to more scalable dynamic approaches that combine concrete and symbolic execution [8, 14, 15, 31, 34]. Dynamic Symbolic Execution (DSE) runs the program under test both concretely—executing the actual program with concrete inputs (e.g., the input string of a parser)—and symbolically, by calculating constraints on the values of program variables expressed in terms of input parameters. In essence, during processing an input the access operations and comparison of input symbols with literals are collected as path conditions.

The basic idea of DSE involves three steps:

1. Running the program with initial or random input,
2. gathering constraints on inputs at conditional statements,
3. and using an SMT (satisfiability modulo theory) [4] solver to generate new program inputs.

This process is repeated and is only constrained by the available time resources. DSE has been successfully implemented in industry-standard tools DART [14], CUTE [31], SAGE [15], PEX [34] and KLEE [7]. Our approach is based on the DSE engine part of the eKnows [25] platform.

The current effectiveness of DSE is limited when applied to programs with highly structured inputs, such as compilers and interpreters [13, 27]. These applications process inputs in stages, including lexing, parsing, and evaluation. In the lexing stage, individual characters are scanned to recognize tokens (terminal symbols of a grammar), such as keywords of programming languages. The subsequent parser employs token-based input grammar checking, in which a token sequence represents a specific case of the input grammar. Inputs generated through symbolic execution at the character level in the lexing stage are usually rejected during the parsing step. Due to the enormous number of control paths in the early stages of processing, DSE rarely reaches parts of the application beyond these initial stages.

In contrast to DSE, pFuzzer [23] is an input generation technique specifically tailored for input parsers. This technique systematically produces inputs for the parser and tracks comparisons made between input characters and expected characters. After each rejection, the comparisons leading to the rejection are satisfied, resulting in new input. pFuzzer only uses the comparison to the last character to create new inputs. This is based on the idea that a parser processes the input character by character and compares these values against possible valid values. The parser then either accepts or rejects these characters. The assumption is that as soon as a character is accepted, everything up to that character is also accepted and can serve as a possible prefix for longer inputs.

In our approach presented in the next section, we use Dynamic Symbolic Execution, integrate pFuzzer's key idea of leveraging comparisons with tokens to generate new inputs, and extend it with a three-phase approach to overcome the path explosion problem for programs with structured inputs.

## 3 Approach

Mimid uses handcrafted inputs in order to generate grammars. Our approach now automatically generates inputs to remove the need for handcrafted inputs. The main idea is to use DSE to generate extended inputs from simpler inputs.

### 3.1 Input Generation using DSE

The problem with DSE in combination with highly structured inputs is the number of paths that have to be explored, leading to the path explosion problem [8]. To reduce this problem, we reuse the idea from pFuzzer [23]. That means, our algorithm uses an input fragment, which has already been recognized being valid, and extends it with a placeholder character, which is intended to be invalid. Executing the program with the extended input gives us path conditions rejecting the placeholder character. Then, an SMT solver is used for creating an input character which is recognized instead of the placeholder symbol. As a placeholder character we use the tilde character '∼'.

The approach is presented in Algorithm 1. It uses a list *inputs* containing already found valid inputs and a process queue *queue* holding the inputs still to be processed.

The algorithm starts with the placeholder character ('∼') as a single input and puts it into the process queue (line 3). Then, in a loop running until the process queue is empty or a time limit is reached, the next input from the process queue is handled (line 5). The program is then run using DSE, which yields whether the input is accepted or rejected by the parser. In addition, the path conditions (PC) from the current run are returned (line 6). If the input is valid, it will be stored in the list of valid inputs (lines 7-9). Next, the PC is filtered for terms containing comparisons to the placeholder character. Let's consider the following input '{∼', which may yield a path condition like in[0] = '{' AND in[1] ≠ '"' AND in[1] ≠ '}' (compare Listing 1 line 20 and 29). Our placeholder character in the input was at index position 1, which means the terms that we are interested in are: in[1] ≠ '"' and in[1] ≠ '}' (line 10). The list of terms is then iterated in a loop (line 11). The term is negated and an SMT solver is used to find a new valid character for the placeholder position (line 12). This placeholder symbol is then replaced by the new character (line 13) and the new input is used to execute the program (line 14). If the input is valid, it will be added to the list of valid inputs (lines 15-17). Then the placeholder character will be appended to the input, and it will be put back into the queue and the process repeats (lines 18-19).

Let's illustrate the algorithm with a simple example: We start with the placeholder character '∼' and execute the program the first time. A PC input[0] ≠ '"' may be

---

**Algorithm 1** Base algorithm for generating new inputs using DSE.

1: *inputs* ← [ ]
2: *queue* ← [ ]
3: *queue.Add*("~")
4: **while not** (*queue.IsEmpty*() **or** *LimitReached*()) **do**
5:     *input* ← *queue.Remove*()
6:     *pathCondition*, *valid* ← *ExecuteProg*(*input*)
7:     **if** *valid* **then**
8:         *inputs.Add*(*input*)
9:     **end if**
10:    *filteredTerms* ← *FilterPC*(*pathCondition*)
11:    **for** *term* **in** *filteredTerms* **do**
12:        *newCh* ← *Solve*(*Not*(*term*))
13:        *newInput* ← *input.ReplaceLastCh*(*newCh*)
14:        *valid* ← *ExecuteProg*(*newInput*)
15:        **if** *valid* **then**
16:            *inputs.Add*(*input*)
17:        **end if**
18:        *newInput* ← *newInput* + "~"
19:        *queue.Add*(*newInput*)
20:    **end for**
21: **end while**

---

returned. Following our approach, we use the PC and generate a new input '''. We append the placeholder character again and repeat our code. After a few steps, the input may look like: '"abc"'. This input is then accepted by the program, and we have successfully generated a first valid input, which represents a JSON string.

### 3.2 Three-Phase Approach Overview

Algorithm 1 reduces the amount of explored paths, however, it fails to discover the deeper paths in parsers that are parsing highly structured inputs. Therefore, a three-phase approach has been introduced to further mitigate the path explosion problem. The main idea is to learn inputs for each grammar rule separately. Mimid creates a grammar rule for each parser function in the code. Thus, the three-phase approach finds valid inputs separately for the individual parser functions.

The inputs for parser functions are divided into three parts: (i) the *function prefix*, which is the input fragment that results in calling the function, (ii) the *function stem*, which is the input fragment recognized by the parser function and (iii) the *function suffix* which represents the rest of a complete valid input.

The three-phase approach now works as follows:

1. In the first phase we are collecting prefixes as well as an initial set of function stems.

**Generating Inputs for Grammar Mining using Dynamic Symbolic Execution**

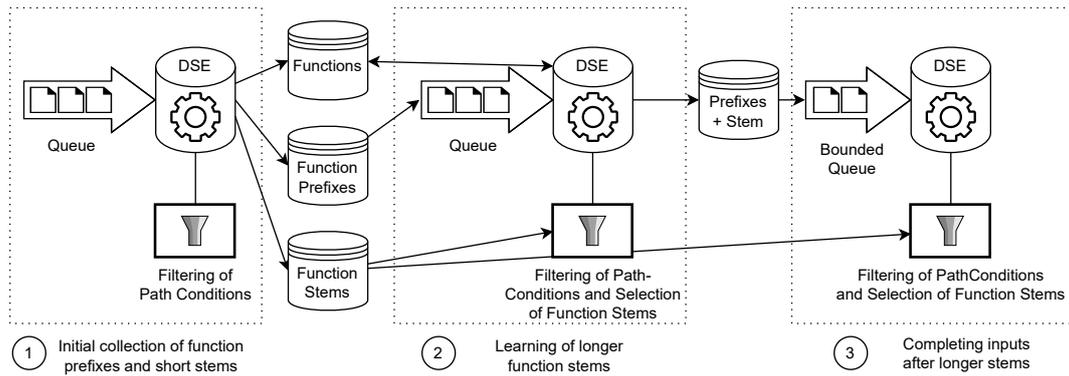

**Figure 2** This figure shows the three-phase approach of the presented concept. (1) The initial collection of the function prefixes and function stems. (2) The usage of the collected function prefixes and generation of longer function stems. (3) The reduction of the queue size for completing inputs after longer function stems.

2. The second phase now uses the prefixes and the initial set of function stems to generate a more complete set of function stems.
3. The third phase extends the input fragments from the first two phases to complete the inputs with function suffixes.

Figure 2 illustrates our approach and shows how the three phases interact with each other. The three phases are marked with rectangular striped boxes, where the data stores in between these boxes depict the information flow between the different steps. The first box integrates Algorithm 1, and the collection of function prefixes and stems. The second one shows the adapted filtering process as well as the generation of longer function stems. In the third box, the limitation of the queue size is depicted leading to the completion of inputs.

The following sections will describe these three phases in detail.

## 3.3 Initialization

During the first phase, the algorithm learns both function prefixes and initial function stems. Figure 3 illustrates how the Microjson program processes an example input, breaking it down into various function prefixes and stems. The figure displays the sample input on the first line, followed by a breakdown of which parts of the input correspond to specific function prefixes and stems. For instance, the function *json_dict* has the prefix '{' and the remainder is the function stem, while *json_string* uses the prefix '{"' and considers the text up to the next '"' as its stem.

### 3.3.1 Collecting Function Prefixes

Algorithm 1 is used to run the program and create the first inputs. During this process, function calls are tracked. Each time a function is entered, the current path conditions can be used to determine the prefix of the function.

Our experiments showed that generating 500 inputs in the first phase is enough to provide prefixes for all parsing functions in our examples. The results of the prefix collection process for the Microjson example are displayed in Table 1. This table contains all the functions that were called. These functions can be split into scanner and parser functions. Typical scanner functions are called for almost every character that occurs in the input. Thus, such functions contain a very high amount of prefixes. Examples for such functions are *len*, *next*, *peek*, *pos*, *substring*. As we are interested in parser functions, we neglect those functions. The interesting functions are the parser-like functions, i.e., functions which "consume" a token. These can be identified with a lower number of found prefixes. In our experiments, we disregard the function if number of prefixes constitute more than 10 % of the tested inputs. The relevant parsing function, for the Microjson example, are functions starting with *json_*.

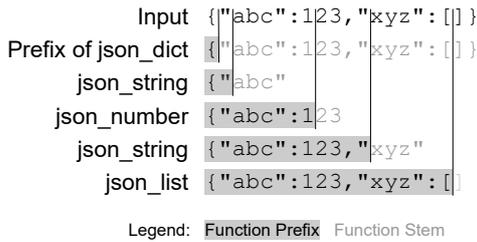

Figure 3 This figure shows an example input, and the function prefixes and function stems that would be collected out of the input.

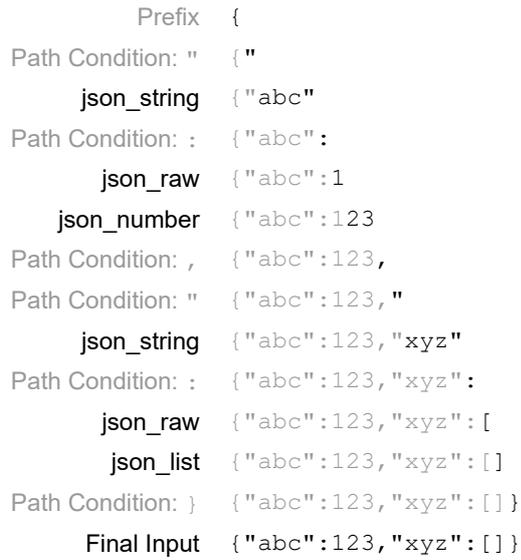

Figure 4 Example which shows how the input string '{"abc":123,"xyz":[]}' is generated by the approach when generating inputs for *json_dict*.

Note, it is the last character sequence of the prefix that is responsible for calling the function. As we see in Table 1 there are multiple prefixes that end with the same character. It would be sufficient to consider one of those. However, in the current version of our implementation we use all of them. This guarantees that all triggering characters are handled.

### 3.3.2 Collecting Initial Function Stems

Besides the function prefixes, function stems are also generated in the initialization phase. When a function prefix is found, we continue adding characters using Algorithm 1. As long as the PC generating the new characters belongs to the same function, the characters are also part of the function stem. A character with which the function is left represents the end of the function stem. The collected function stems (shown in Table 2) represent the initial set of function stems used in the next phase. Note that

**Generating Inputs for Grammar Mining using Dynamic Symbolic Execution**

■ **Table 1** This table shows an excerpt of function prefixes that were collected from the Microjson program. The values in the prefix column are separated by a semicolon followed by a space.

| Function | Prefix | Number of prefixes |
| --- | --- | --- |
| _decode_utf8 | "~~; ["~; {"~; [["~; … | 7 |
| decode_escape | "\ | 1 |
| getvalue | [1~; 9.~; 4E~; … | 534 |
| json_dict | [{; [[{; { | 3 |
| json_fixed | [n; t; [[t; f; n; | 9 |
| json_list | [; [[[; [ | 3 |
| json_number | [[+; [[4; +; .; 3; … | 39 |
| json_raw | [{; [[; [4; ["; … | 42 |
| json_string | "; [{"; [["; ["; {" | 5 |
| len | [1~; 9.~; "; +; … | 453 |
| next | 00; 01; 02; "; +; ["~; [{"~; … | 262 |
| next2 | true; false; true; null | 4 |
| peek | 00; [[4~; 01; 1+; … | 299 |
| pos | [1~; 9.~; "; +; … | 525 |
| read | 00; 0E; 1.; "; -; .; {"~; … | 262 |
| read2 | [true; null; false; true | 4 |
| skipspaces | [{; [[4~; [~; … | 70 |
| substr | [1~; 9.~; 4E~; "; -; .; … | 532 |

the number of functions for which function stems are generated is smaller than the number of functions associated with prefixes. The reason is that functions exist which do not parse an input character and therefore have no function stem. New inputs are added at the back of the queue, due to the first in first out principle. This leads to shorter inputs being prioritized thus resulting in shorter function stems.

### 3.4 Generating Longer Function Stems

The initialization phase provides a set of prefixes and an initial collection of function stems for each function. In this second phase, we try to generate longer function stems. We start with a function prefix and use the approach from Algorithm 1 to extend the prefix. However, when a subfunction is called we do not extend the input for the subfunction but instead insert a function stem from the set of function stems of that subfunction. Thus, we reuse function stems rather than generating new stems.

Figure 4 illustrates the approach by an example based on the *json_dict* method from Listing 1. We start with the prefix '{' and add the placeholder symbol: '{~'. In this case, we are in the *json_dict* function that we want to learn, and we proceed with using the path conditions from that function. The path condition may yield that a possible character for our placeholder position is '"'. This means that in the next iteration the

■ **Table 2** This table shows an excerpt of the function stems that are accepted per function. The function stems shown in the table are separated by a semicolon followed by a space.

| Function | Valid input | Number of valid inputs |
| --- | --- | --- |
| json_dict | } | 1 |
| from_json | 88; oo; 89; … | 156 |
| json_fixed | alse; rue; ull | 3 |
| json_list | ] | 1 |
| json_number | 0; 1; 2; 3; … | 11 |
| json_raw | 88; oo; 89; … | 156 |
| json_string | " | 1 |
| skipspaces | "; f; o; .; [; ]; … | 21 |

input '{"~' is present. In this iteration, the execution returns that the last function that has processed a character was *json_string*. Thus, we check if the current value (without our placeholder) is a valid prefix for the function *json_string* and if there exists a function stem for *json_string*. In this case, a valid function stem is 'abc"'. This is added to the input, which leaves the *json_string* function and returns to *json_dict*. We continue with Algorithm 1 and find ':' as the next valid character. Following our approach, we generate a new character '1' which leads to the *json_raw* call. As the *json_raw* function dispatches the '1' to the *json_number* function, a function stem for *json_number* has to be found. A valid function stem is '23', which we append to the input. This process is then iteratively continued, until finally, in the *json_dict* method we find '}', which then results in the final valid input of e.g., '{"abc":123,"xyz":[]}'.

This process is implemented in Algorithm 2, which represents an extension of Algorithm 1. The algorithm uses the function prefixes and stems collected in the previous phase as an initial input. It then starts by iterating over all functions for which prefixes were collected (line 1). These prefixes together with the placeholder character (~) are then inserted into the queue (lines 3-5). Next, we continue with the loop from the base algorithm as long as the queue is not empty or a time limit is reached (line 6). After that, the program is executed with the input resulting in the PC, the flag that represents whether the input is valid, and the name of the function that last accessed the placeholder character (line 8). If the input is valid, it will then be added to the list of valid inputs (lines 9-11).

Next, we initialize a flag to indicate whether processing via the PC approach is required. This flag is later used to determine whether to proceed with the PC approach, either when analyzing the current function to be learned or when no function stem is found (line 12). Following that, a check is applied if the current function is the function to learn or if it is a call to another function (lines 13-14). If the latter is the case, a function stem for the function that last accessed the placeholder character is extracted (line 15). If a stem is found (line 16), the placeholder character at the end of the input is replaced (line 17), and the function stem together with the placeholder

Generating Inputs for Grammar Mining using Dynamic Symbolic Execution

---

**Algorithm 2** Adapted version of Algorithm 1, that includes both the collected prefixes and valid function values. The base algorithm is displayed gray, whilst changes are shown in black.

---

  **Given:** collectedPrefixes (*colPre*): A dictionary of the collected function prefixes
  **Given:** collectedFuncStems (*colStem*): A dictionary of the collected function stems
  **Given:** inputs: The set of already collected valid inputs

1: **for** *funcToLearn, prefixes* **in** *colPre* **do**
2:  *queue* ← [ ]
3:  **for** *prefix* **in** *prefixes* **do**
4:   *queue*.Add(*prefix* + " ∼ ")
5:  **end for**
6:  **while not** (*queue*.IsEmpty() **or** $LimitReached()$) **do**
7:   *input* ← *queue*.Remove()
8:   *pathCondition, valid, lAccFunc* ← $ExecuteProg(input)$
9:   **if** *valid* **then**
10:    *inputs*.Add(*input*)
11:   **end if**
12:   *needPC* ← *True*
13:   *funcStart* ← $isFuncStart(colPre, lAccFunc, input)$
14:   **if** *funcToLearn* ≠ *lAccFunc* **and** *funcStart* **then**
15:    *validValue* ← *colStem*.GetForFunc(*lAccFunc*)
16:    **if** *validValue* **then**
17:     *newInput* ← *input*.removeLastChar()
18:     *newInput* ← *newInput* + *validValue* + " ∼ "
19:     *queue*.Add(*newInput*)
20:     *needPC* ← *False*
21:    **end if**
22:   **end if**
23:   **if** *needPC* **then**
24:    *filteredTerms* ← $FilterPC(pathCondition, lAccFunc)$
25:    **for** *pc* **in** *filteredTerms* **do**
26:     …
27:    **end for**
28:   **end if**
29:  **end while**
30: **end for**

---

character is appended (line 18). Then the input is added to the queue (line 19). In this case, no further processing using PC is needed, and the *needPC* flag is set to false.

If further processing via PC is necessary (line 23), because we are either in the function that we currently want to learn inputs for or no function stem was found, we continue with the base algorithm (lines 24-28). The only modification made to the base algorithm is that we additionally filter PC to only contain terms that were created in the function last accessing the placeholder character (line 24). This helps to further reduce the path explosion and allows exploring the search space more directed to the function for which we want to learn inputs. If during this process new functions with prefixes are found, this process repeats, until no new functions are found anymore.

We conducted experiments on the evaluated examples with time budgets ranging from 5 to 500 seconds per function before transitioning to the final phase. However, increasing the budget beyond 30 seconds does not improve the final result.

### 3.5 Completing Inputs

In phase 2 we were able to generate inputs with function prefixes and longer function stems. Nevertheless, there is still the problem that inputs may be still incomplete and must be completed for valid inputs. Until this phase, we had an unbounded queue size. To ensure that inputs are being completed, the queue size is limited. The limited queue size causes the inputs in the queue to get processed more often and thus have a greater chance of being completed.

Currently, we select a random subset of inputs from the queue. We found out that selecting them purely randomized gives the best diversity and thus leads to the best results. Our experiments with different queue sizes showed that reducing to a size of 50 elements works out best for the examples that are contained in the evaluation.

Except for the change in queue size, the algorithm remains the same as in phase 2, meaning that suffixes are found by iteratively expanding the input. This final phase runs until all 50 inputs are accepted by the program or a 10-second time limit is reached, which experiments show works well for the evaluated examples.

## 4 Evaluation

We implemented our input generation approach in Java based on the eKnows DSE engine [25]. We also implemented the Mimid approach in Java, based on the algorithms given in [17]. This allowed us to do the whole evaluation in Java. A replication package is provided under [2].

The objective of our evaluation is to assess the quality of the inputs generated by our approach, with a particular focus on demonstrating the effectiveness of the proposed three-phase strategy. It is important to emphasize that the primary aim is not to enhance the grammar mining process itself, but rather to facilitate grammar mining by automatically generating suitable inputs, thereby eliminating the need for manually crafted ones. To this end, we compare the grammars derived from inputs generated under four distinct evaluation settings:

**Generating Inputs for Grammar Mining using Dynamic Symbolic Execution**

1. **Handwritten Inputs:** This setting serves as a reference for the performance of our grammar mining approach using handwritten or existing inputs from the field. While other grammar miner implementations may achieve higher scores, our focus is on input generation, not the miner. Note that existing inputs may not cover all features, and our reimplementation of Mimid may differ slightly from the original due to some minor implementation details.
2. **Klee:** As a representative of state-of-the-art symbolic execution engines, we include a comparison using Klee [7]. This setting evaluates how a general-purpose DSE tool performs in generating inputs for grammar mining, thereby providing insight into the suitability of classic DSE techniques for this task.
3. **One-Phase Approach:** To isolate the impact of our full three-phase approach, we first evaluate the performance when only the initial phase is applied. This corresponds to the algorithm presented in Algorithm 1, which reuses the idea of pFuzzer [23]. It should be noted, however, that unlike the original pFuzzer implementation, our version does not incorporate the same set of heuristics, and is thus not directly comparable in terms of performance.
4. **Three-Phase Approach:** Finally, we evaluate the complete three-phase input generation approach proposed in this work. This setting is intended to demonstrate the practical benefits and overall effectiveness of the full strategy in supporting grammar mining tasks.

In order to have comparable numbers, we measure the accuracy of the grammar as a producer (*precision*) and the accuracy of the derived grammar as parsers (*recall*), as also done in [17, 33]. The two metrics are calculated as follows:

*Precision* is measured by checking how many inputs generated by the learned grammar are also accepted by the program. High precision implies that the language of the learned grammar closely matches what the program accepts.

*Recall* is measured by checking how many of the inputs that the program accepts are also accepted by the grammar. High recall indicates that the learned grammar has a broad coverage of what the program considers valid. As it is practically not possible to know the inputs that a program accepts, a so-called golden grammar is used, which perfectly represents the input space that the program accepts. These golden grammars were handcrafted and checked by multiple experts to ensure correctness.

To summarize, our research method consists of four steps:

1. input creation using the four described methods,
2. grammar mining with the method of Mimid using the inputs,
3. using the mined grammar as generator for inputs and measuring precision, i.e., if the inputs are accepted by the program,
4. using the mined grammar as parser, i.e., using the golden grammar for generating inputs and testing if the mined grammar accepts the inputs (recall).

Most of the grammar mining approaches define their own set of benchmark programs. Our work is based on the results of the Mimid approach, thus we decided to primarily focus on their benchmark suite. Our current approach only allows analyzing Python programs, thus we decided to include the Python programs (*calc.py*,

*mathexpr.py*, *cgidecode.py* and *microjson.py*) from their work. These programs are of moderate complexity, containing only a few parser methods. Therefore, we manually translated the larger C program *mjs.c*, which includes 33 parser methods, into Python to better demonstrate the effectiveness of our approach.

Additionally, we included the examples from FbGL [33] as they provided the grammars in their replication package and their parsers strictly follows the recursive descent pattern, allowing us to manually translate the programs (*AdvExpr*, *ExprParser*, *HelloParser*, *JsonParser*, *MailParser* and *UrlParser*) to Python. For evaluation scenario A, handwritten inputs are required. For the Mimid examples, we use the inputs that are available in their replication package. For the FbGL examples, the authors handcrafted the inputs.

To enable a fair comparision between the different methods it is necessary to ensure equal evaluation settings. Thus, we generated inputs using our three phase approach with the settings described in Section 3. We then counted the number of *valid* generated inputs, and used that as a goal for the other two settings. The complexity of the programs is shown in Table 4. In addition, the number of generated inputs, plus the average and maximum length of the inputs is shown in Table 6.

In the one-phase setting, only the first phase of our described algorithm is executed. The only difference is that other configurations are applied. Instead of switching to phase two after generating 500 inputs, this phase is executed until the desired number of valid inputs or a time limit of one hour is reached. If the latter one is the case, we stop the execution and work with the inputs, that we have until then, or if no inputs are found, we define the problem as unsolvable by the algorithm.

In order to construct our evaluation example with Klee, a few steps were necessary. First, the Python programs had to be translated to C. This step was performed manually by the one of authors, and checked by the other authors ensuring correctness. The code was translated as close to the Python implementation as possible, to ensure comparability. The time limit for Klee was adjusted individually for each program to ensure that at least the desired number of inputs were generated. If no valid inputs were found within one hour, the program was considered unsolvable by Klee.

## 5 Results

Based on the experimental setup outlined in Section 4, we collected results for each of the four evaluation settings. Table 3 summarizes the measured precision, recall, and the corresponding F1-score. The scenario using manually crafted inputs serves as a reference point for our evaluation in Table 5.

Before interpreting the results, we briefly clarify the metrics used. A high precision indicates that the grammar predominantly generates syntactically valid inputs—however, even a trivial grammar that generates only a single valid input could yield a precision of 100 %. Conversely, a high recall suggests that the grammar accepts a large proportion of the inputs accepted by the target program—but this could also be achieved by a grammar that accepts all possible strings, including invalid ones. Therefore, precision and recall must always be considered in conjunction.

Generating Inputs for Grammar Mining using Dynamic Symbolic Execution

■ **Table 3** Results of the conducted experiments, which show the precision (Prec) and recall (Rec) as well as the F1-Score the three automatic approaches: Klee, One-Phase and our Three-Phase Approach. All scores are show in percentage.

|  | Klee | | | One-Phase | | | Three-Phase | | |
| --- | --- | --- | --- | --- | --- | --- | --- | --- | --- |
| Program | Prec | Rec | F1 | Prec | Rec | F1 | Prec | Rec | F1 |
| **AdvExprParser** | 100.0 | 100.0 | 100.0 | 100.0 | 100.0 | 100.0 | 100.0 | 100.0 | 100.0 |
| **ExprParser** | 100.0 | 100.0 | 100.0 | 100.0 | 100.0 | 100.0 | 100.0 | 100.0 | 100.0 |
| **HelloParser** | 100.0 | 100.0 | 100.0 | 100.0 | 100.0 | 100.0 | 100.0 | 100.0 | 100.0 |
| **Calc** | 100.0 | 100.0 | 100.0 | 100.0 | 8.0 | 14.8 | 100.0 | 100.0 | 100.0 |
| **MailParser** | 0.0 | 0.0 | 0.0 | 0.0 | 0.0 | 0.0 | 100.0 | 100.0 | 100.0 |
| **UrlParser** | 0.0 | 0.0 | 0.0 | 0.0 | 0.0 | 0.0 | 100.0 | 96.5 | 98.2 |
| **CgiDecode** | 100.0 | 33.3 | 50.0 | 96.9 | 96.9 | 96.9 | 97.7 | 97.3 | 97.5 |
| **JsonParser** | 100.0 | 97.2 | 98.6 | 100.0 | 84.2 | 91.4 | 100.0 | 88.6 | 94.0 |
| **Mjs** | 100.0 | 21.8 | 35.8 | 100.0 | 37.8 | 54.9 | 87.5 | 77.2 | 82.0 |
| **MicroJson** | 99.1 | 56.7 | 72.1 | 70.2 | 70.5 | 70.3 | 74.7 | 71.8 | 73.2 |
| **Mathexpr** | 100.0 | 58.0 | 73.4 | 100.0 | 50.5 | 67.1 | 100.0 | 25.4 | 40.5 |
| **Average** | 81.7 | 60.6 | 66.4 | 78.8 | 58.9 | 63.2 | 96.4 | 87.0 | 89.6 |

To this end, we report and interpret the F1-score as the primary performance indicator, as it balances both precision and recall into a single metric, providing a more meaningful measure of overall grammar quality.

The programs AdvExprParser, ExprParser, and HelloParser achieved perfect scores across all approaches. Each of these programs contains only a small number of parsing methods (6, 4, and 2, respectively) that strictly follow the principles of recursive descent parsing. In addition, they perform character-by-character comparisons only, without checking for multi-character sequences such as keywords. Combined with a reduced set of digits (1–3 instead of 0–9), this simplicity allows all approaches to generate sufficient inputs, resulting in perfect scores.

The Calc program shows a significant improvement in recall and F1 score with the Three-Phase approach compared to the One-Phase approach (100 % vs. 8 % recall). Although the program uses few parsing methods only, the One-Phase approach struggles with one parsing method that performs stateful parsing based on a flag to track operators. As a result, no inputs containing parentheses were generated, leading to the low recall score.

MailParser and UrlParser show a significant improvement with the Three-Phase approach, achieving near-perfect scores compared to Klee and One-Phase, which fail entirely. Both Klee and the One-Phase approach struggle to terminate generated sequences of letters (a–z) due to path explosion. In the case of MailParser, the generated input consists solely of letters and never reaches the @ character. Consequently, both approaches fail to generate a single valid input.

For CgiDecode, the Three-Phase approach shows a clear improvement over Klee, which generates only a single random character in places where arbitrary characters except % are allowed. As a result, the grammar miner struggles to generalize these characters into an ANY symbol. Compared to the One-Phase approach, however, the improvement is minimal, since the parser relies on a single parser method, obviously leaving little room for the Three-Phase approach to provide additional benefit.

For JsonParser, the Three-Phase approach achieves a high score, but shows no improvement over KLEE. Manual inspection of the extracted grammar reveals that no inputs are generated containing multiple digits after a decimal point, which seems that our approach preferred the wrong path conditions in this phase. Compared to the One-Phase approach, the Three-Phase approach performs slightly better, as the inputs lack deeper structures and repetition.

For Mjs, the Three-Phase approach shows a good improvement, primarily due to the complexity of the program, which includes many parser methods and a complex structure. These factors make it more challenging for the other two approaches.

For MicroJson, all three approaches perform nearly equal. Since parsing is not done in many methods, the Three-Phase approach cannot significantly outperform the other two. Compared to KLEE, it achieves better recall, but this comes at the cost of worse precision, which balances out the overall performance.

In MathExpr, our approach falls short compared to the other two. The main reason is that none of the approaches effectively detect the mathematical functions and variables within the parser. However, the other two approaches manage to detect one more variable than ours, which may be due to our approach not utilizing all path conditions in the final phases, thus not being able to find all those structures. This is currently considered a known limitation of our approach.

A comparison between our Three-Phase approach and the manually crafted inputs (as shown in Table 5) shows that both methods achieve perfect results for the first five test cases. In four of the remaining six examples, the manually crafted inputs capture certain input aspects more completely than our approach. However, in the cases of CgiDecode and MicroJson, which reused the Mimid inputs, our Three-Phase method was able to identify additional structural details that the manual inputs missed, leading to better grammar mining results.

Table 6 presents the average and maximum input lengths generated by each approach. It shows that our Three-Phase approach produces significantly longer inputs, which can help uncover subtle features that might be missed by shorter inputs. Additionally, Table 7 provides runtime measurements for both input generation and grammar mining, allowing a more complete assessment of overall performance. The results indicate that grammar mining is the most time-consuming step, while input generation contributes only a small portion. However, due to the longer inputs, the Three-Phase approach requires more time for both generation and the mining process.

■ **Listing 3** Excerpt of the golden grammar proposed in Mimid for the math expression example.

```
1 mathexpr_golden = {
2   "<word>": [ "acos", "sin", ... ],
3   "<factor>": [ "<word>(<expr>,<expr>)", "<word>", ... ], ...
4 }
```

## 5.1 Answer to Research Questions

This section presents answers to our research questions based on our results.

### 5.1.1 Answer to RQ 1

The presented approach successfully generates meaningful inputs for grammar mining. By incrementally expanding inputs through dynamic symbolic execution and leveraging a three-phase strategy, the method ensures that the generated inputs are both relevant and comprehensive.

### 5.1.2 Answer to RQ 2

Table 3 shows that our approach performs relatively well compared to other input generation techniques. In particular, the comparison with KLEE highlights that our method is often more effective at uncovering deeper structures in more complex programs. Additionally, the comparison with the One-Phase approach demonstrates the effectiveness of our Three-Phase strategy in generating high-quality inputs.

# 6 Discussion

The discovered results are dependent on various factors, some of which are beyond our full control.

One factor influencing the results is randomization. To ensure consistency, we used seeded random values. While fuzzing inherently involves randomness, we mitigated its impact by repeating the evaluation multiple times, which consistently yielded the same results. Although it's difficult to guarantee identical test inputs across all approaches, we minimized this effect by using the same fuzzer as Mimid.

The quality of the golden grammars also impacts evaluation. Although they were crafted and reviewed by experts, inaccuracies remain. For example, Gopinath et al.'s grammar [17] does not fully reflect actual parser behavior. In the math expression case (Listing 3), functions like *sin* and *cos* are defined to take two or zero parameters, making valid single-parameter calls invalid. Since only inputs accepted by the program are generated, such mismatches can inflate recall by excluding unsupported constructs.

After potential challenges to the study's rigor have been addressed, a comprehensive discussion of the results follows, considering their broader implications and insights.

While the numeric results show strong performance across most examples, they do not fully capture a core strength of our approach: the ability of generated inputs to uncover hidden edge cases. Unlike handcrafted inputs, which may overlook certain structures even with expert knowledge, generated inputs systematically explore the input space. For example, our method identified multiple commas in JSON dictionaries, a subtle case missed in existing inputs as discussed in Section 2. This advantage becomes particularly clear in the result for CgiDecode and MicroJson cases, where generated inputs exposed gaps in the resulting grammars. In this way, automated input generation not only reduces manual effort but also leads to more complete and accurate grammar extraction.

## 7 Related Work

Our core idea, extending dynamic symbolic execution (DSE), addresses the mitigation of the path explosion problem when dealing with programs with structured inputs. In this section, we compare our approach with related extensions of DSE and other grammar mining approaches.

Pygmalion [16] extends pFuzzer by learning an input grammar from comparisons made between input characters and using grammar-based fuzzing for test generation. It tracks the data flow of input characters during program execution, similar to the Autogram [18] approach. Pygmalion was only evaluated for fuzzing by measuring the number of valid inputs compared to the total number of inputs generated, which is essentially our precision evaluation score. For the comparable programs *MathExpr* and *Microjson* (referred to as Json in their work), Pygmalion achieved precision scores of 73.6 % and 77.8 %, respectively. Our approach closely matches these results, with precision scores of 100 % for *MathExpr* and 74.7 % for *Microjson*.

REINAM [36], integrates dynamic symbolic execution for finding a few initial input samples and then continues with reinforcement learning. They evaluate their approach on a different set of examples. However, we can compare their *mailto* example with our *9) MailParser* example. REINAM achieves approximately 93 % precision and 77 % recall, while our approach reaches 100 % for both metrics.

Pan et al. [27] propose a symbolic execution approach that can automatically generate token sequences to test complex parsing programs. Their key idea is to symbolize tokens instead of input characters to improve the efficiency of symbolic execution. While this method facilitates the generation of program inputs that pass the lexer step, it does not provide further guidance to reduce the complexity of constraints collected during the parsing step. This is necessary for grammar mining and is addressed by our three-phase approach.

Godefroid et al. [13] propose treating lexer output as symbolic input and generating constraints at the token level rather than the character level. Constraints are first collected at the character level, then used to construct tokens with corresponding constraints. Unlike traditional DSE approaches, new input is generated using a path condition and input grammar. However, since this method requires an existing input grammar, it is not suitable for grammar mining.

Bettscheider and Zeller [6] extract input grammars by transforming the program under test, abstracting parsing functions into symbolic nonterminals and restricting loops and recursion to ensure a finite exploration space. They apply symbolic execution to systematically explore parsing paths without relying on any input samples. In contrast, our approach leverages dynamic symbolic execution (DSE) within a structured three-phase input generation strategy to generate inputs for the dynamic grammar miner Mimid and to incrementally expand the input space.

Kim et al. [19] present an approach for the automatic extraction of grammars for fuzzing. They use a concolic testing approach and track API-level function calls to extract string comparisons, thereby extracting grammar-like structures. This approach does not require any inputs to extract the grammar. However, due to the limitations of concolic testing, such as branch explosion, this approach seems unsuitable for deeper parser structures.

Godefroid [12] extends DART with SMART, which uses function summaries, including preconditions and postconditions, to reuse results and speed up testing. Similarly, Anand et al. [1] propose a demand-driven compositional symbolic execution that builds test inputs by composing symbolic function summaries, avoiding full path exploration. They use logic formulas and SMT solvers to scale symbolic execution. In contrast, our approach learns specific functions using function prefixes and focuses not on outputs, but on identifying accepted input sets.

Additionally, there is ongoing work from Schröder and Cito [30] where they describe a concept on how to transform programs into intermediate representations that make the control flow explicit and use that information to inference grammars. At their current state of work, it is not clear yet, what the exact requirements for the inputs are, but due to the fact, that they also want to analyze the control flow, we expect them to require a set of valid inputs as well.

## 8  Conclusion and Future Work

In this work, we presented an approach to enhance existing grammar miners through automatic input generation. We introduced a three-phase input generation approach that first collects function prefixes and valid function inputs, then generates inputs targeting specific functions. This significantly reduces complexity and ensures proper input coverage for each function. Finally, the algorithm closes the inputs by limiting the processing queue size. We demonstrated our input generation on the grammar miner Mimid, showing that this approach is effective for grammar mining and that generated inputs perform as well as, or better than, handwritten ones.

The results revealed that parsers frequently check not only for syntactical correctness but also consider semantic constraints. This was particularly evident in the *MathExpr* parser, where mathematical functions often restrict parameter values to specific ranges. Such information cannot be captured by basic formal grammars. Therefore, in our future work, we aim to enhance grammar mining by incorporating the extraction of semantic information from parser systems. Initial concepts for extracting semantics have been explored by Mera [24], Moser et al. [26], Kovačević et al. [21] and Pointner [29]. We believe that incorporating attribute grammars can lead to significant improvements in fuzz testing.

**Acknowledgements**   The research reported in this work has been partly funded by BMIMI, BMWET, and the State of Upper Austria in the frame of the SCCH competence center INTEGRATE (FFG grant no. 892418) part of the FFG COMET Competence Centers for Excellent Technologies Programme. We also gratefully acknowledge funding from the Austrian Research Promotion Agency (FFG) and the University of Applied Sciences Upper Austria as part of the project AG-Fuzzer (FFG grant no. 895972).

## 9 Appendix

■ **Table 4** This table presents the complexity of each program using two metrics: the number of parser methods and lines of code.

| Program | Nr. of Parser Methods | Lines of Code |
|---|---|---|
| **AdvExprParser** | 6 | 71 |
| **Calc** | 3 | 40 |
| **CgiDecode** | 1 | 27 |
| **ExprParser** | 4 | 57 |
| **HelloParser** | 2 | 44 |
| **JsonParser** | 21 | 197 |
| **MailParser** | 5 | 42 |
| **Mathexpr** | 9 | 213 |
| **MicroJson** | 9 | 218 |
| **Mjs** | 33 | 822 |
| **UrlParser** | 20 | 291 |

■ **Table 5** Results of the conducted experiments, which show the precision (as producer) and recall (as parser) as well as the F1-Score for the handwritten inputs.

| Program | Precision | Recall | F1 |
|---|---|---|---|
| **AdvExprParser** | 100.0 | 100.0 | 100.0 |
| **ExprParser** | 100.0 | 100.0 | 100.0 |
| **HelloParser** | 100.0 | 100.0 | 100.0 |
| **Calc** | 100.0 | 100.0 | 100.0 |
| **MailParser** | 100.0 | 100.0 | 100.0 |
| **UrlParser** | 100.0 | 100.0 | 100.0 |
| **CgiDecode** | 99.9 | 72.8 | 84.2 |
| **JsonParser** | 100.0 | 100.0 | 100.0 |
| **Mjs** | 88.5 | 100.0 | 93.9 |
| **MicroJson** | 100.0 | 43.5 | 60.6 |
| **Mathexpr** | 87.3 | 67.0 | 75.8 |
| **Average** | **97.8** | **89.4** | **92.2** |

■ **Table 6** This table shows the number of inputs generated by our Three-Phase approach, which serves as the target input count for the other two approaches. Addtionally, it compares the average and maximum input length of the three input generation approaches.

|  |  | Klee |  | One-Phase |  | Three-Phase |  |
|---|---|---|---|---|---|---|---|
| Program | Nr. of Inputs | Avg. | Max | Avg. | Max | Avg. | Max |
| **AdvExprParser** | 600 | 4.2 | 6 | 4.2 | 5 | 34.2 | 67 |
| **Calc** | 473 | 6.4 | 15 | 3.8 | 4 | 28.0 | 85 |
| **CgiDecode** | 100 | 4.5 | 7 | 4.4 | 5 | 4.7 | 5 |
| **ExprParser** | 337 | 5.8 | 7 | 4.7 | 5 | 16.7 | 37 |
| **HelloParser** | 32 | 6.0 | 6 | 6.0 | 6 | 6.0 | 6 |
| **JsonParser** | 586 | 12.5 | 31 | 8.4 | 8 | 33.6 | 147 |
| **MailParser** | 240 |  |  |  |  | 28.2 | 54 |
| **Mathexpr** | 802 | 5.7 | 9 | 3.8 | 4 | 27.1 | 124 |
| **MicroJson** | 1290 | 7.0 | 9 | 3.9 | 6 | 18.5 | 53 |
| **Mjs** | 3365 | 5.9 | 20 | 3.8 | 17 | 24.1 | 241 |
| **UrlParser** | 1868 |  |  |  |  | 41.7 | 54 |

■ **Table 7** This table shows the runtime (in ms) of the Input Generation (IG) and Grammar Mining (GM) of the three tested approaches. For Klee the Input Generation was defined with a time limit, thus no detailed time measurements are reported.

|  | Klee | One-Phase |  | Three-Phase |  |
|---|---|---|---|---|---|
| Program | GM | IG | GM | IG | GM |
| **AdvExprParser** | 11 760 | 1 980 | 13 509 | 20 208 | 82 486 |
| **Calc** | 9 195 | 690 | 7 059 | 13 625 | 30 859 |
| **CgiDecode** | 1 063 | 5 389 | 1 366 | 2 914 | 1 218 |
| **ExprParser** | 5 130 | 606 | 5 877 | 3 778 | 9 040 |
| **HelloParser** | 356 | 84 | 711 | 100 | 971 |
| **JsonParser** | 36 958 | 18 405 | 34 394 | 47 075 | 114 256 |
| **MailParser** |  | 3 600 125 |  | 7 165 | 47 444 |
| **Mathexpr** | 71 776 | 9 286 | 51 382 | 54 760 | 234 454 |
| **MicroJson** | 691 044 | 7 235 | 122 951 | 69 274 | 426 067 |
| **Mjs** | 1 094 070 | 23 202 | 845 228 | 280 058 | 2 088 693 |
| **UrlParser** |  | 3 605 090 |  | 70 355 | 744 396 |

## About the authors

**Andreas Pointner** Andreas Pointner is a researcher in the AIST research group at the University of Applied Sciences Upper Austria, Campus Hagenberg. He received both his bachelor's and master's degrees in Software Engineering from the same university and is currently pursuing as a PhD student at Johannes Kepler University Linz, Austria. Contact him at andreas.pointner@fh-hagenberg.at.
https://orcid.org/0000-0001-8642-1161

**Josef Pichler** Josef Pichler is a professor for programming and project development at the University of Applied Sciences in Upper Austria, Campus Hagenberg. His research interests include static code analysis, reverse engineering, and software maintenance. Pichler received his PhD in computer science from the Johannes Kepler University Linz, Austria. Contact him at josef.pichler@fh-hagenberg.at.
https://orcid.org/0009-0007-6908-0474

**Herbert Prähofer** Herbert Prähofer is an Associate Professor at the Institute for System Software at the Johannes Kepler University Linz, Austria. His research interests include software development methods and tools, model-based software development, simulation theory and methods, and software engineering methods in the automation domain. He received a master's degree and a doctoral degree in computer science from the Johannes Kepler University Linz, Austria in 1986 and 1991, respectively. Contact him at herbert.praehofer@jku.at.
https://orcid.org/0000-0002-0139-8044